\pdfoutput=1
\RequirePackage{ifpdf}
\ifpdf 
\documentclass[pdftex]{sigma}
\else
\documentclass{sigma}
\fi

\numberwithin{equation}{section}

\newtheorem*{Theorem*}{Theorem}

\theoremstyle{definition}

\usepackage{tikz-cd}

\def\be{\begin{equation}}
\def\ee{\end{equation}}
\def\bes{\begin{subequations}}
\def\ees{\end{subequations}}

\def\vep{\varepsilon}
\def\i{{\rm i}}
\def\e{{\rm e}}
\def\d{{\rm d}}
\def\ul{\underline}

\def\q{{\mathfrak{q}}}
\def\t{{\mathfrak{t}}}
\def\p{{\mathfrak{p}}}
\def\L{{\text{L}}}

\newcommand{\mb}[1]{\mathbf{#1}}

\def\Xint#1{\mathchoice
 {\XXint\displaystyle\textstyle{#1}}%
 {\XXint\textstyle\scriptstyle{#1}}%
 {\XXint\scriptstyle\scriptscriptstyle{#1}}%
 {\XXint\scriptscriptstyle\scriptscriptstyle{#1}}%
 \!\int}
\def\XXint#1#2#3{{\setbox0=\hbox{$#1{#2#3}{\int}$}
 \vcenter{\hbox{$#2#3$}}\kern-.5\wd0}}

\def\dashint{\Xint-}

\def\ket#1{{ |#1\rangle}}
\def\bra#1{{ \langle#1|}}
\def\braket #1#2{{ \langle #1|#2 \rangle}}

\def\kett#1{{ |#1\rangle\!\rangle}}
\def\braa#1{{ \langle\!\langle#1|}}

\begin{document}

\allowdisplaybreaks

\newcommand{\arXivNumber}{2507.02831}

\renewcommand{\PaperNumber}{003}

\FirstPageHeading

\ShortArticleName{Trace Formulas for Deformed W-Algebras}

\ArticleName{Trace Formulas for Deformed W-Algebras}

\Author{Fabrizio NIERI~$^{\rm ab}$}

\AuthorNameForHeading{F.~Nieri}

\Address{$^{\rm a)}$~Dipartimento di Matematica ``Giuseppe Peano'', Universit\`a di Torino,\\
\hphantom{$^{\rm a)}$}~Via Carlo Alberto 10, 10123 Torino, Italy}
\Address{$^{\rm b)}$~INFN, Sezione di Torino, Via Pietro Giuria 1, 10125 Torino, Italy}
\EmailD{\mail{fb.nieri@gmail.com}}
\URLaddressD{\url{https://sites.google.com/unito.it/mathphysturin/people/nieri}}

\ArticleDates{Received September 08, 2025, in final form January 04, 2026; Published online January 13, 2026}

\Abstract{We investigate trace formulas in $\varepsilon$-deformed W-algebras, highlighting a novel connection to the modular double of $\mathfrak{q}$-deformed W-algebras. In particular, we show that torus correlators in the additive (Yangian) setting reproduce sphere correlators in the trigonometric setup, possibly with the inclusion of a non-perturbative completion. From a dual perspective, this mechanism implements a gauge theoretic 2d$\to$3d uplift, where a~circle direction in the world-sheet transmutes to a compact space-time direction in a non-trivial manner. We further discuss a unified picture of deformed W-algebras driven by trace formulas, suggesting a deeper algebraic layer related to the massive and massless form-factor approach to integrable QFT and 2d CFT.}


\Classification{81T20; 81T40; 81T60; 81R10; 81R50; 17B68}

\section{Introduction}

In this letter, we study quiver ${\rm W}_{\vep_1,\vep_2}$ algebras introduced in \cite{Nieri:2019mdl}, which can be viewed as the additive (or Yangian) counterpart of quiver ${\rm W}_{\q_1,\q_2}$ algebras \cite{Kimura:2015rgi}. The former lie at the bottom of the rational/trigonometric/elliptic classification borrowed from integrable systems, the top layer ${\rm W}_{\q_1,\q_2,\q'}$ being also known \cite{Feigin_2009,Kimura:2016dys,Nieri:2015dts}. While it is fairly straightforward to follow the hierarchy from the elliptic to rational case by taking appropriate limits, the natural question arises whether the lower layers also capture some aspects of the upper ones. Such a possibility is partially hinted by the representation theory of affine Yangian, quantum toroidal and elliptic algebras, which exhibit striking similarities.\footnote{In the Yangian case the free boson representation is more subtle and the S-automorphism is apparently lost, even though it reverberates in certain observables \cite{Jeong:2024onv,Mironov:2012ba,Nieri:2019mdl}.}

We approach the question by exploiting the continuous free boson representation of ${\rm W}_{\vep_1,\vep_2}$ algebras and computing some observables to be matched across the hierarchy, specifically at the trigonometric level. Which type of observables is naturally suggested by the dual string/gauge theoretic realization of the relevant algebras in the BPS sector of 4d/5d/6d theories -- including co-dimension 2 defects -- with 8 supercharges on the $\Omega$-background times a point/circle/torus~\cite{Nekrasov:2015wsu}. For instance, S-duality in IIB string theory implies the equivalence between partition functions of 5d linear quivers with unitary gauge groups and adjoint matter and 6d theories with fundamentals: geometrically, both the Calabi--Yau backgrounds involve a compact direction~-- either due to the adjoint or the torus -- and are simply related via the fiber-base exchange \cite{Hollowood:2003cv}. From the algebraic viewpoint, the identification of certain correlators -- interpreted as gauge theoretic partition functions -- between the trigonometric and elliptic setups can be traced back to Miki S-automorphism or, in practice, to Clavelli--Shapiro trace technique: this trick allows torus correlators~-- i.e., traces over the entire free boson Fock space -- to be recasted as sphere correlators~-- i.e., vacuum expectation values~-- at the cost of doubling the number of bosons, an operation which effectively implements the elliptic deformation. This mechanism was already discussed in \cite{Kimura:2015rgi,Nieri:2019mdl} to connect trigonometric observables/5d instanton partition functions with the elliptic/6d ones: the algebraic or world-sheet circle direction~-- i.e., the trace -- transmutes to an emergent compact space-time direction.

\subsection*{Main result}

This web of relationships naturally suggests that the trigonometric $\q$-deformation can be realized by looking at torus correlators within the $\vep$-deformation. From the gauge theoretic perspective, this operation should be dual to the dimensional uplift 2d$\to$3d.\footnote{We will often trade 4d/5d/6d theories for their 2d/3d/4d vortex defects to simplify the discussion. From the algebraic viewpoint, the main difference regards the types of representations involved.} In the following, we argue this connection does indeed take place in an interesting manner: not only the trace implements the desired jump, but it does so in a way that automatically generates a non-perturbative completion in the sense of the modular double \cite{Nedelin:2016gwu}. In other words, granted that the 2-point functions of the ${\rm W}_{\vep_1,\vep_2}$ screening currents $\mb{S}(X)$ are the main building blocks to construct (integrated) correlators/gauge theoretic partition functions \cite{Aganagic:2013tta,Aganagic:2014oia}, namely
\[
\langle \mb{S}(X)\mb{S}\big(X'\big)\rangle= \text{1-loop determinants on $\mathbb{C}$},
\]
then we find
\[
\operatorname{Tr}\bigl( \e^{-\tau\mb{L}(0)} \mb{S}(X)\mb{S}\big(X'\big)\bigr)= \text{1-loop determinants on $\mathbb{S}^3$}.
\]
In previous works, such compact result stemmed from a fusion \cite{Nieri:2013yra} echoing left/right factorization in 2d CFT or topological/anti-topological factorization in gauge theories on compact spaces~\cite{Beem:2012mb,Pasquetti:2011fj}.
From the viewpoint of integrated correlators, projecting to a ``chiral'' $\mathbb{C}\times\mathbb{S}^1$ factor\footnote{We recall that $\mathbb{S}^3$ admits the genus 1 Heegaard splitting into two solid tori $\mathbb{C}\times\mathbb{S}^1$.} amounts to choosing the ``perturbative'' contour, which neglects the ``non-perturbative'' poles.\footnote{It is well-known that 1-loop determinants of 3d $\mathcal{N}=2$ gauge theories on $\mathbb{S}^3$ can be written in terms of the Double Sine function, which has two towers of simple poles at discrete points along two directions in the complex plane, usually denoted by $b$, $1/b$. It is customary to refer to the poles along $b$ as perturbative while those along~$1/b$ as non-perturbative. See also \cite{Pasquetti:2011fj} for a Borel resummation analysis.}

\subsection*{Further motivations and connections to other works}
\begin{enumerate}\itemsep=0pt
\item[$(1)$] The main motivation behind this letter is the idea that trigonometric and elliptic W-algebras should emerge explicitly from massive integrable QFTs, just as ordinary W-algebras arise from (massless) 2d CFTs. In this regard, the original AGT correspondence~\cite{Alday:2009aq,Wyllard:2009hg} goes far beyond a mere identification of symmetries: it relates a specific class of 4d $\mathcal{N}=2$ gauge theories \cite{Gaiotto:2009we} to specific 2d conformal models (Liouville/Toda). In the deformed setup, substantial progress has been achieved on the gauge theoretic side, but less so on the other: while the representation theory of quantum toroidal algebras is now well-developed and inspired by gauge/string theory constructions, very little is known about the associated integrable QFTs.\footnote{The gauge/Bethe side of the story \cite{Nekrasov:2009uh} is more developed, but we do not touch it here.} Nevertheless, the natural theoretical framework to study the latter can be found in the form-factor approach \cite{doi:10.1142/1115}. In particular, Lukyanov~\cite{Lukyanov:1993pn} developed bosonization techniques that express form-factors through trace formulas -- a~language well suited to the present work.\footnote{Already in the early papers on the subject many formulas similar to ours can be found, see, e.g., \cite{Jimbo:1996ev} and \cite{Lukyanov:1995gs}.} We hope that contextualizing the deformed W-algebras within the form-factor program may shed more light on the relevant massive models dual to higher-dimensional gauge theories.

\item[$(2)$] A renewed interest in the massless form-factor program \cite{Delfino:1994ea,Fendley:1993jh,ZAMOLODCHIKOV1992602} and the realization of affine Yangian/${\rm W}_{\vep_1,\vep_2}$ algebras in this context, may trigger some progress on long-standing open issues in 2d CFT, such as the computation of Toda 3-point functions -- \cite{Coman:2019eex,Isachenkov:2014eya,Mitev:2014isa} for a string/gauge theoretic approach -- or holographic problems \cite{Bielli:2023gke,Torrielli:2021hnd}. We provide further comments in this direction in the last section.

\item[$(3)$] When computing integrated correlators, one can in principle consider all poles, namely also those captured by the non-perturbative contour. Since certain ${\rm W}_{\q_1,\q_2}$ correlators are dual to open string amplitudes, it would be interesting to consider our findings in view of various proposals for a non-perturbative definition of (refined) topological strings \cite{Hatsuda:2013oxa,Lockhart:2012vp}.

\item[$(4)$] On the more algebraic side, the realization of 3d $\mathcal{N}=2$ holomorphic blocks as torus correlators in ${\rm W}_{\vep_1,\vep_2}$ may also help explaining their quantum modular properties \cite{Cheng:2022rqr,Cheng:2018vpl}. Furthermore, the ${\rm W}_{\vep_1,\vep_2}$/${\rm W}_{\q_1,\q_2}$ connection may clarify why the (refined) topological vertex can be understood both as a quantum toroidal intertwiner \cite{Awata:2011ce} (trigonometric viewpoint) and a VOA character \cite{Prochazka:2015deb} (affine Yangian viewpoint).

\end{enumerate}

\noindent The rest of this letter is organized as follows. In Section \ref{sec:conventions}, we briefly review some definitions and conventions around ${\rm W}_{\q_1,\q_2}$ and ${\rm W}_{\vep_1,\vep_2}$ algebras, at least for the simplest $A_1$ case. In Section~\ref{sec:trace}, we compute the trace of the product of two screening currents of the ${\rm W}_{\vep_1,\vep_2}(A_1)$ algebra, matching the 2-point function in the modular double completion of ${\rm W}_{\q_1,\q_2}(A_1)$. In Section \ref{sec:discussion}, we comment further our results and outline directions for future work. In Appendix \ref{app:SF}, we summarize the relevant special functions. In Appendix \ref{app:CS}, we give a proof of Clavelli--Shapiro formula.

\section{Brief review and conventions}\label{sec:conventions}

\subsection*{The $\q$-Virasoro algebra}

For the sake of simplicity, we focus here on the ${\rm W}_{\q,\t^{-1}}(A_1)$ algebra, namely $\q$-Virasoro.\footnote{It is customary to set $\q_1=\q$ and $\q_2=\t^{-1}$ to match the refined topological string literature.} The generalization to arbitrary quiver W-algebras is straightforward.\footnote{Following, e.g., \cite{Kimura:2015rgi}, one has to introduce an index for each node and a non-trivial deformed Cartan matrix, not necessarily associated to a Lie algebra, encoding the structure of the quiver. The extension to fractional quivers is also possible following \cite{Kimura:2017hez}.} In order to start with, let us consider the Heisenberg algebra generated by oscillators $\{\mb{a}_m, m \in \mathbb{Z}\backslash \{0\} \}$ and zero modes~${\{\mb{P}, \mb{Q}\}}$, with non-trivial commutation relations
\[
 \bigl[\mb{a}_m, \mb{a}_n\bigr] = -\frac{1}{m}\bigl(\q^{m/2} - \q^{-m/2}\bigr)\bigl(\t^{-m/2} - \t^{m/2}\bigr)C^{[m]}(\p)\delta_{m+n, 0} , \qquad \bigl[\mb{P}, \mb{Q} \bigr] = 2 ,
\]
where $C^{[m]}(\p)=\bigl(\p^{m/2} + \p^{-m/2}\bigr)$ is the deformed Cartan matrix of $A_1$ and $\p\equiv \q\t^{-1}$. For any given $\alpha\in\mathbb{C}$, we consider dual Fock modules over the charged Fock vacua \smash{$\ket{\alpha}\equiv \e^{\alpha\mb{Q}/2}\ket{0}$} and~${\bra{\alpha}\equiv\bra{0}\e^{-\alpha\mb{Q}/2}}$ with the canonical pairing $\braket{0}{0}=1$. In particular,
\[
\mb{P}\ket{\alpha}=\alpha\ket{\alpha} , \qquad \mb{a}_m\ket{\alpha}=0 ,\qquad \bra{\alpha}\mb{a}_{-m}=0 ,\qquad m\in\mathbb{Z}_{>0}.
\]

The $\q$-Virasoro screening current\footnote{There is a second screening current related to this one by $\q\leftrightarrow \t^{-1}$.} has the following free boson representation:
\[
 \mb{S}(x) \equiv :\e^{\mb{s}(x)}:,\qquad \mb{s}(x)\equiv - \sum_{m \ne 0} \frac{\mb{a}_m x^{-m}}{\q^{m/2} - \q^{- m/2}} +\sqrt{\beta}\mb{Q}+\sqrt{\beta}\mb{P}\ln x,
\]
where $\t\equiv\q^\beta$. The normal ordering symbol $:~:$ arranges all the positive modes and $\mb{P}$ to the right. For definitiveness, in the following we shall consider the chamber $|\q|<1$, then the OPE of the screening current with itself is
\be\label{eq:SSqOPE}
\mb{S}(x)\mb{S}(x')= :\mb{S}(x)\mb{S}(x'): \frac{(x'/x;\q)_\infty (\p x'/x;\q)_\infty}{(\q x'/x;\q)_\infty(\t x'/x;\q)_\infty}\, x^{2\beta}.
\ee
The corresponding 2-point function is the main building block to construct non-trivial (integral) correlators of suitably chosen vertex operators, therefore this is the object we study in this letter.

\subsection*{The $\boldsymbol{\vep}$-Virasoro algebra}\label{sec:vepAlg}

The $\vep$-Virasoro algebra can be thought of as the additive counterpart of $\q$-Virasoro. The former can be deduced from the latter by a careful scaling limit, part of which consists in parametrizing
\[
\q\equiv \e^{-\hbar \vep_1},\qquad \t\equiv \e^{\hbar \vep_2},\qquad \p\equiv \e^{-\hbar \vep_+},\qquad x\equiv \e^{\hbar X},\qquad \vep_+\equiv\vep_1+\vep_2,
\]
and taking $\hbar \to 0$. Then the screening current\footnote{There is another one with $\vep_1$ and $\vep_2$ exchanged which we do not consider here.} takes the free boson representation (we use the same symbol)
\[
\mb{S}(X) \equiv :\e^{\mb{s}(X)}:,\qquad \mb{s}(X)\equiv \dashint \d k \frac{\mb{a}(k)\, \e^{-k X}}{2\sinh(\vep_1 k/2)}+\sqrt{\beta}\mb{P}X,
\]
where the oscillator modes, now labelled by a continuous parameter $k\in\mathbb{R}$, satisfy the non-trivial commutation relations
\[
\big[\mb{a}(k),\mb{a}\big(k'\big)\big]=-\frac{4}{k}\sinh(\vep_1k/2)\sinh(\vep_2k/2)\delta\big(k+k'\big)C(k),
\]
with $C(k)=2\cosh(\vep_+ k/2)$. The Fock modules and normal ordering are defined as before, while the dashed integral means the contribution from around the origin (zero mode) is to be regularized (e.g., via principal value or some subtraction scheme). The OPE of the screening current with itself then reads as
 \[
\mb{S}(X)\mb{S}\big(X'\big)= :\mb{S}(X)\mb{S}\big(X'\big): \frac{\Gamma_1\big(X-X'-\vep_2|\vep_1\big)\Gamma_1\big(X-X'+\vep_1|\vep_1\big)}{\Gamma_1\big(X-X'|\vep_1\big)\Gamma_1\big(X-X'+\vep_+|\vep_1\big)} \e^{-\gamma \vep_1\vep_2}.
\]

\section[W\_q\_1,q\_2 sphere correlators as W\_vep\_1,vep\_2 torus correlators]{$\boldsymbol{{\rm W}_{\q_1,\q_2}}$ sphere correlators as $\boldsymbol{{\rm W}_{\vep_1,\vep_2}}$ torus correlators}\label{sec:trace}

We move from 2-point functions, i.e., v.e.v.\ or sphere correlators, to torus correlators. We shall adapt Clavelli--Shapiro trace technique to the continuous setup needed for the free boson representation of ${\rm W}_{\vep_1,\vep_2}$ algebras. Let us review their formula for the discrete case first (we refer to Appendix \ref{app:CS} for a proof and further details). Let us introduce a grading operator $\mb{L}_0$ which satisfies
$
[\mb{L}_0,\mb{a}_{n}]=-n\, \mb{a}_n$,
and define the torus correlator
$
\braa{0}\mb{O}\kett{0}\equiv \operatorname{Tr}\bigl( \q'^{\L_0}\mb{O}\bigr)$,
where $\mb{O}$ is some operator written in terms of the non-zero modes of the Heisenberg algebra and $\q'\in\mathbb{C}^\times$ the elliptic parameter. The trace is taken over the entire Fock space. The Clavelli--Shapiro trace technique tells us that we can equivalently compute\footnote{The proportionality factor turns out to be $1/\bigl(\q';\q'\bigr)_\infty$.}
$
\braa{0}\mb{O}\kett{0}\simeq \bra{0}\hat{\mb{O}}\ket{0}$,
where $\hat{\mb{O}}$ is the elliptic version of the operator $\mb{O}$, namely the former is obtained from the latter through the substitution
\[
\mb{a}_n\to \frac{\mb{a}_n}{1-\q'^n}+\mb{b}_{-n},\qquad \mb{a}_{-n}\to \mb{a}_{-n}-\frac{\mb{b}_{n}}{1-\q'^{-n}},\qquad n\in\mathbb{Z}_{>0},
\]
where the oscillators $\mb{a}_{n}$, $\mb{b}_{n}$ satisfy the same Heisenberg algebra and commute with each other. Similarly, in the continuous case we define a grading operator such that\footnote{This can be thought of as the scaling limit $\hbar\mb{L}_0\to \mb{L}(0)$ for $\hbar \to 0$.}
$
[\mb{L}(0),\mb{a}(k)]=-k\, \mb{a}(k)$,
namely
\[
\mb{L}(0)\equiv -\dashint_0^\infty\d k \frac{k^2\,\mb{a}(-k)\mb{a}(k)}{4\sinh(k\vep_1/2)\sinh(k\vep_2/2)C(k)}.
\]
We now fix $\tau\in\mathbb{C}$ and aim to compute traces weighted by \smash{$\e^{-\tau\mb{L}(0)}$}. Applying the same strategy as before, we get the formula
$
\braa{0}\mb{O}\kett{0}\simeq \bra{0}\hat{\mb{O}}\ket{0}$,
where on the right-hand side the following substitutions are implied
\[
\mb{a}(k)\to \frac{\mb{a}(k)}{1-\e^{-k\tau}}+\mb{b}(-k),\qquad \mb{a}(-k)\to \mb{a}(-k)-\frac{\mb{b}(k)}{1-\e^{k\tau}},\qquad k\in\mathbb{R}_{>0}.
\]
As before, the $\mb{a}(k)$, $\mb{b}(k)$ operators satisfy the same algebra and commute with each other. Some regularization scheme is understood and discussed in the following.

Let us now apply this result to compute the weighted trace of two $\vep$-Virasoro screening currents (up to zero modes), namely
\begin{gather*}
\bra{0}\hat{\mb{S}}(X)\hat{\mb{S}}\bigl(X'\bigr)\ket{0}\\
\qquad= \exp\left[\dashint_{0}^{+\infty}\frac{\d k}{k} \frac{\e^{-k(X-X'+\vep_1/2)}\bigl(\e^{k\vep_2/2}-\e^{-k\vep_2/2}\bigr)\bigl(\e^{k\vep_+/2}+\e^{-k\vep_+/2}\bigr)}{\bigl(1-\e^{-k\vep_1}\bigr)\bigl(1-\e^{-k\tau}\bigr)}\right]\\
\phantom{\qquad=}{}\times\exp\left[-\dashint_{0}^{+\infty}\frac{\d k}{k} \frac{\e^{-k(\tau-X+X'-\vep_1/2)}\bigl(\e^{k\vep_2/2}-\e^{-k\vep_2/2}\bigr)\bigl(\e^{k\vep_+/2}+\e^{-k\vep_+/2}\bigr)}{\bigl(1-\e^{k\vep_1}\bigr)\bigl(1-\e^{-k\tau}\bigr)}\right].
\end{gather*}
Now it comes the key observation. If we adopt the Hankel regularization separately for the two integrals as outlined below \eqref{eq:GammaInt}, we don't seem to get something sensible to be interpreted in the \mbox{$\q$-Virasoro} context. Indeed, using the integral representation (\ref{eq:GammaInt}), we would naively expect the appearance of a certain combination of Double Gamma functions which would nicely combine into the ratio (\ref{eq:SSqOPE}) of $\q$-Pochhammers (up to zero mode contributions) thanks to the identity (\ref{eq:GGqP}). However, the validity of (\ref{eq:GammaInt}) would require $\tau$ and $\pm\vep_1$ to all lie on the same side with respect to the imaginary axis, which cannot be the case. Interestingly, the case at hand is a rare example where there is in fact another regularization, simpler to understand. We first combine the two integrals into a single one by performing the change of variable $k\to - k$ in the second piece, so that
\[
\bra{0}\hat{\mb{S}}(X)\hat{\mb{S}}\big(X'\big)\ket{0}
=\exp\left[\dashint_{-\infty}^{+\infty}\frac{\d k}{k} \frac{\e^{-k(X-X'+\vep_1/2)}\bigl(\e^{k\vep_2/2}-\e^{-k\vep_2/2}\bigr)\bigl(\e^{k\vep_+/2}+\e^{-k\vep_+/2}\bigr)}{\bigl(1-\e^{-k\vep_1}\bigr)\bigl(1-\e^{-k\tau}\bigr)}\right].
\]
The dashed integration can now be naturally taken along the entire real axis with a small deformation around the origin in order to avoid it (equivalently, the contour can run parallel to the real axis slightly below/above). Using the integral representation (\ref{eq:Srint}), the expression above is the ``compact version'' of the expected naive result (up to zero mode contributions), namely
\be\label{eq:qWcompact}
\bra{0}\hat{\mb{S}}(X)\hat{\mb{S}}\big(X'\big)\ket{0}=
\frac{S_2(X-X'|\tau,\vep_1)S_2(X-X'+\vep_+|\tau,\vep_1)}{S_2(X-X'+\vep_1|\tau,\vep_1)S_2(X-X'-\vep_2|\tau,\vep_1)} \e^{\pm \frac{2\pi\i\beta}{\tau} (X-X'+\tau/2)},
\ee
where the $\q$-Pochhammers are replaced by double sine functions.

\section{Discussion and outlook}\label{sec:discussion}

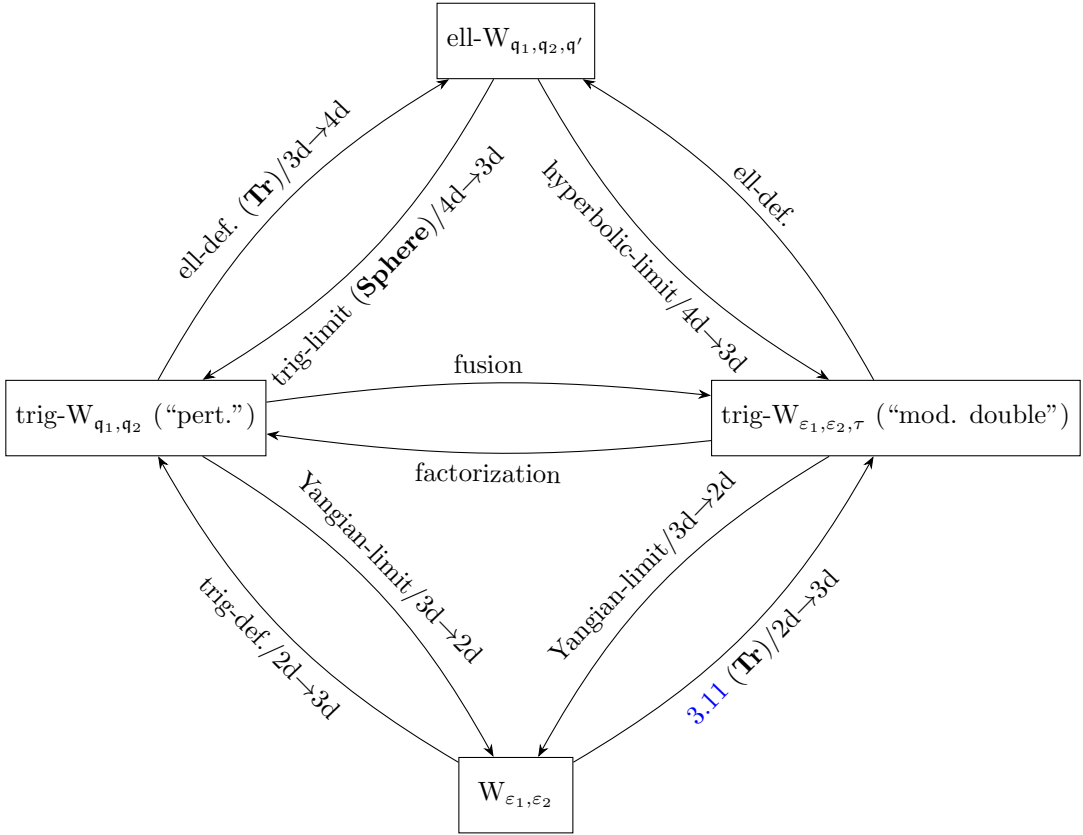
\begin{figure}[ht!]
\centering
\begin{tikzpicture}[>=Stealth, font=\small, node distance=4cm]

 \node[draw, rectangle, minimum width=1.5cm, minimum height=1cm] (top) at (0, 5) {$\text{ell-W}_{\q_1,\q_2,\q'}$};
 \node[draw, rectangle, minimum width=1.5cm, minimum height=1cm] (bottom) at (0, -5) {${\rm W}_{\vep_1,\vep_2}$};
 \node[draw, rectangle, minimum width=1.5cm, minimum height=1cm] (left) at (-5, 0) {$\text{trig-W}_{\q_1,\q_2}$ (``pert.")};
 \node[draw, rectangle, minimum width=1.5cm, minimum height=1cm] (right) at (5, 0) {$\text{trig-W}_{\vep_1,\vep_2,\tau}$ (``mod. double")};

 \draw[<-, bend left=15] (right) to node[sloped, below] {hyperbolic-limit/4d$\to$3d} (top);
 \draw[<-, bend left=15] (top) to node[sloped, above] {ell-def.} (right);

 \draw[<-, bend left=15] (bottom) to node[sloped, above] {Yangian-limit/3d$\to$2d} (right);
 \draw[<-, bend left=15] (right) to node[sloped, below] {\ref{eq:qWcompact} (\textbf{Tr})/2d$\to$3d} (bottom);

 \draw[->, bend left=15] (bottom) to node[sloped, below] {trig-def./2d$\to$3d} (left);
 \draw[->, bend left=15] (left) to node[sloped, above] {Yangian-limit/3d$\to$2d} (bottom);

 \draw[->, bend left=15] (left) to node[sloped, above] {ell-def. (\textbf{Tr})/3d$\to$4d} (top);
 \draw[->, bend left=15] (top) to node[sloped, below] {trig-limit (\textbf{Sphere})/4d$\to$3d} (left);

 \draw[->, bend left=7] (left) to node[sloped, above] {fusion} (right);
 \draw[->, bend left=7] (right) to node[sloped, below] {factorization} (left);

\end{tikzpicture}
\caption{Sketch of the rational/trigonometric/elliptic classification of the deformed W-algebras and the related limits and uplifts one can implement on gauge theoretic observables (for simplicity, at co-dimension 2 in 4d/5d/6d) and the involved special functions or observables.}\label{fig:hier}
\end{figure}

\noindent The results above can be placed at the bottom-right corner of the cartoon in Figure~\ref{fig:hier}, which provides a schematic overview of the rational/trigonometric/elliptic classification of deformed W-algebras and their associated gauge-theoretic backgrounds, along with the interrelations realized via appropriate limits and uplifts.
\begin{enumerate}\itemsep=0pt
\item[$(1)$] At the top level, we encounter the (master) elliptic deformation, which admits two limiting cases, named trigonometric and hyperbolic. In gauge theoretic terms, the former is associated to the $\mathbb{C}_{\q_1} \times \mathbb{S}^1$ background, the latter to the compact space \smash{$\mathbb{S}^3_{\vep_1/\tau}$}. This structure is mirrored in the corresponding 2-point functions/1-loop determinants, which are expressed in terms of $\q$-Pochhammers or double sine functions respectively. From a~practical perspective, both cases are accessible because of the automorphic properties of the key building block -- the elliptic Gamma function.

\item[$(2)$] At the intermediate level, the trigonometric deformation can be uplifted to the elliptic one~\cite{Nieri:2015dts}, corresponding to the $\mathbb{C}_{\q_1}\times\mathbb{T}^2_{\q'}$ background. The hyperbolic deformation maps to~the modular double upgrade (see, e.g., \cite{Nedelin:2016gwu} for an extensive discussion of that construction), whereas the elliptic uplift replaces \smash{$\mathbb{S}^3_{\vep_1/\tau}$} with \smash{$\mathbb{S}^3_{\vep_1/\tau}\times\mathbb{S}^1$} on the gauge theoretic side (see, e.g., \cite{Lodin:2017lrc}). In any case, the Yangian/additive deformation can be reached by scaling limits.

\item[$(3)$] At the bottom level, the Yangian/additive deformation was derived in \cite{Nieri:2019mdl} by implementing the gauge theoretic dimensional reduction on the algebraic side. In this letter, we went backward and showed the trace implements the hyperbolic deformation or, equivalently, the 2d$\to$3d uplift in the compact sense -- cf.\ bottom-right part of Figure~\ref{fig:hier}. This interpretation is further supported by the observation that trace manipulations with continuous free bosons require a regularization scheme, the simplest being a discretization of the modes~\cite{Lukyanov:1993pn} -- cf.\ bottom-left part of Figure~\ref{fig:hier}. Subsequently, the continuous limit applied to torus/elliptic correlators -- cf.\ top-right part of Figure~\ref{fig:hier} -- precisely yields the hyperbolic deformation, thus completing the circle.
\end{enumerate}

As we mentioned in the introduction, contextualizing the deformed W-algebras within the form-factor program may shed some light on the massive integrable QFTs dual to higher-dimensional gauge theories. This letter is only a first step in this direction. Another intriguing research line is to explore whether the cartoon in Figure~\ref{fig:hier} conceals an additional deeper layer below the $\vep$-deformation. In this case, the observables of the Yangian deformation should in turn arise from trace formulas associated with a more fundamental algebra. This would establish a~direct connection with the massless form factor approach to 2d CFT -- a possibility we intend to investigate elsewhere.

\appendix

\section{Multiple Gamma and sine functions}\label{app:SF}
We follow the reference \cite{naru}. For $z,\omega_i\in\mathbb{C}$ all lying on the same side with respect to the imaginary axis of the complex plane, the multiple Gamma function has the following integral representation:
\be\label{eq:GammaInt}
\Gamma_r(z|\ul\omega)=\exp\left[\frac{\gamma}{r!}(-1)^r B_{rr}(z|\ul\omega)+\oint_\mathcal{C}\frac{\d k}{2\pi\i k}\ln(-k)\frac{\e^{-k z}}{\prod_{j=1}^r\bigl(1-\e^{-k \omega_j}\bigr)}\right],
\ee
where $\mathcal{C}$ is the Hankel contour, $\gamma$ the Euler--Mascheroni constant and $B_{rr}$ the multiple Bernoulli polynomials. This type of representation is useful whenever one has to deal with divergent integrals extending from $0$ to $\infty$ as the substitution
\[
\int_0^\infty \d k \cdots \to \oint_\mathcal{C}\frac{\d k}{2\pi\i k}\ln(-k) \cdots
\]
yields a sensible regularization around the origin. See, e.g., \cite{Nieri:2019mdl} and references therein for more details.

There is an identity \cite{FRIEDMAN2004362} converting the product of two multiple Gamma functions into a~multiple $\q$-Pochhammer, namely
\begin{gather}
\Gamma_r(x|1,\ul\alpha)
\Gamma_r(1-x|1,-\ul\alpha)=\frac{\e^{-\frac{\i\pi}{r!}(-1)^r B_{rr}(x|1,\ul\alpha)}}{(x;\ul\alpha)_\infty},\nonumber\\ (x;\ul\alpha)_\infty\equiv \bigl(\e^{2\pi\i x};\e^{2\pi\i\alpha_1},\dots, \e^{2\pi\i\alpha_{r-1}}\bigr)_\infty.\label{eq:GGqP}
\end{gather}

The multiple sine function for $\operatorname{Re} \omega_j>0$, $\operatorname{Re}\sum_j\omega_j>\operatorname{Re}z>0$ has a similar integral representation
\be\label{eq:Srint}
S_r(z|\ul \omega)=\exp\left[\pm(-1)^r\frac{\pi \i}{r!}B_{rr}(z|\ul\omega)-\oint_{\mathbb{R}\mp \i 0}\frac{\d k}{k}\frac{\e^{-k z}}{\prod_{j=1}^r\bigl(1-\e^{-k\omega_j}\bigr)}\right],
\ee
where the contour runs parallel to the real axis avoiding the origin.

\section{Clavelli--Shapiro trace technique}\label{app:CS}

We offer a step-by-step proof of Clavelli--Shapiro trace formula, which cannot be easily found in the literature. We aim to compute
$
Z=\operatorname{Tr} x^{\mb{L_0}} \mb{O}\bigl(\mb{a},\mb{a}^\dagger\bigr)$,
where $\mb{O}$ is some operator-valued function of $\mb{a}$, $\mb{a}^\dagger$, the usual annihilation/creation operators, and $\mb{L}_0\equiv\mb{a}^\dagger \mb{a}$ is the occupation number operator. They satisfy the non-trivial commutation relations
$
\bigl[\mb{a},\mb{a}^\dagger\bigr]=1$, $ [\mb{L}_0,\mb{a}]=-\mb{a}$, $ \bigl[\mb{L}_0,\mb{a}^\dagger\bigr]=\mb{a}^\dagger$.
The Fock modules (over which we compute the trace) are generated by the vacua~$\bra{0}$,~$\ket{0}$ such that
$
\mb{a}\ket{0}=\bra{0}\mb{a}^\dagger=0$,
with $\braket{0}{0}=1$. Normalized basis states are given by
\[
\ket{n}=\frac{\bigl(\mb{a}^\dagger\bigr)^n}{\sqrt{n!}}\ket{0},\qquad \bra{n}=\bra{0}\frac{(\mb{a})^n}{\sqrt{n!}},\qquad n\in\mathbb{Z}_{\geq 0}.
\]
We start noticing that $Z$ can be rewritten as a vacuum expectation value rather than as a trace by introducing another set of oscillators $\mb{b}$, $\mb{b}^\dagger$ satisfying the very same commutation relations and commuting with $\mb{a}$, $\mb{a}^\dagger$. Indeed
\begin{align*}
\bra{0}\e^{\mb{a}\mb{b}}x^{\mb{L}_0}\mb{O}\bigl(\mb{a},\mb{a}^\dagger\bigr)\e^{\mb{a}^\dagger\mb{b}^\dagger}\ket{0}={}& \sum_{n,m\geq 0}\bra{0}\frac{\mb{a}^n\mb{b}^n}{n!} x^{\mb{L}_0} \mb{O}\bigl(\mb{a},\mb{a}^\dagger\bigr)\frac{\bigl(\mb{a}^\dagger\bigr)^m\bigl(\mb{b}^\dagger\bigr)^m}{m!}\ket{0}\\
= {}& \sum_{n,m\geq 0}\bra{0}\frac{\mb{a}^n}{n!} x^{\mb{L}_0} \mb{O}\bigl(\mb{a},\mb{a}^\dagger\bigr)\frac{\bigl(\mb{a}^\dagger\bigr)^m}{m!}\ket{0}\bra{0}\mb{b}^n\bigl(\mb{b}^\dagger\bigr)^m\ket{0}\\
= {}& \sum_{n,m\geq 0}\bra{0}\frac{\mb{a}^n}{n!} x^{\mb{L}_0} \mb{O}\bigl(\mb{a},\mb{a}^\dagger\bigr)\frac{\bigl(\mb{a}^\dagger\bigr)^m}{m!}\ket{0} m!\delta_{m,n}\\
={} & \sum_{n\geq 0}\bra{n} x^{\mb{L}_0} \mb{O}\bigl(\mb{a},\mb{a}^\dagger\bigr)\ket{n}.
\end{align*}
We must now compute the left-hand side in an alternative way by exploiting ordering and disentangling identities. Firstly, due to $\bigl[\mb{a}^\dagger\mb{b}^\dagger,\mb{a}\bigr]=-\mb{b}^\dagger$, $\bigl[\mb{a}^\dagger\mb{b}^\dagger,\mb{a}^\dagger\bigr]=\bigl[\mb{a}^\dagger\mb{b}^\dagger,\mb{b}^\dagger\bigr]=0$, the BCH formula implies
\smash{$
\e^{-\mb{a}^\dagger\mb{b}^\dagger}\mb{a}\e^{\mb{a}^\dagger\mb{b}^\dagger}=\mb{a}+\mb{b}^\dagger$}, \smash{$ \e^{-\mb{a}^\dagger\mb{b}^\dagger}\mb{a}^\dagger\e^{\mb{a}^\dagger\mb{b}^\dagger}=\mb{a}^\dagger$},
and therefore
\[
\bra{0}\e^{\mb{a}\mb{b}}x^{\mb{L}_0} \mb{O}\bigl(\mb{a},\mb{a}^\dagger\bigr)\e^{\mb{a}^\dagger\mb{b}^\dagger}\ket{0}=\bra{0}\e^{\mb{a}\mb{b}}x^{\mb{L}_0} \e^{\mb{a}^\dagger\mb{b}^\dagger}\mb{O}\bigl(\mb{a}+\mb{b}^\dagger,\mb{a}^\dagger\bigr)\ket{0}.
\]
Now the fact that $\mb{L}_0$ is the grading/dilation operator for the $\mb{a}$, $\mb{a}^\dagger$ modes (or the BCH formula again) implies
\smash{$
x^{-\mb{L}_0}\e^{\mb{a}\mb{b}}x^{\mb{L}_0}=\e^{x\mb{a}\mb{b}}$},
so
\[
\bra{0}\e^{\mb{a}\mb{b}}x^{\mb{L}_0} \e^{\mb{a}^\dagger\mb{b}^\dagger}\mb{O}\bigl(\mb{a}+\mb{b}^\dagger,\mb{a}^\dagger\bigr)\ket{0}=\bra{0}\e^{x\mb{a}\mb{b}} \e^{\mb{a}^\dagger\mb{b}^\dagger}\mb{O}\bigl(\mb{a}+\mb{b}^\dagger,\mb{a}^\dagger\bigr)\ket{0}.
\]
Using the commutation rule $\bigl[-\mb{a}^\dagger\mb{b}^\dagger,x \mb{a}\mb{b}\bigr]=x\bigl(\mb{a}\mb{a}^\dagger+\mb{b}^\dagger \mb{b}\bigr)$ and the BCH formula once more, namely
\smash{$
\e^{-\mb{a}^\dagger\mb{b}^\dagger}\e^{x\mb{a}\mb{b}}\e^{\mb{a}^\dagger\mb{b}^\dagger}=\e^{x(\mb{a}\mb{b}+\mb{a}\mb{a}^\dagger+\mb{b}^\dagger \mb{b}+\mb{a}^\dagger \mb{b}^\dagger)}$},
we get
\[
\bra{0}\e^{x\mb{a}\mb{b}} \e^{\mb{a}^\dagger\mb{b}^\dagger}\mb{O}\bigl(\mb{a}+\mb{b}^\dagger,\mb{a}^\dagger\bigr)\ket{0}=\bra{0}\e^{x(\mb{a}+\mb{b}^\dagger)(\mb{a}^\dagger+\mb{b})} \mb{O}\bigl(\mb{a}+\mb{b}^\dagger,\mb{a}^\dagger\bigr)\ket{0}.
\]
In order to evaluate the exponential operator on the vacuum, we can follow at least two routes. The elegant one is based on the observation that $\mb{a}^\dagger\mb{b}^\dagger\equiv \mb{E}$, $\mb{a}\mb{b}\equiv \mb{F}$, $[\mb{E},\mb{F}]=\mb{a}^\dagger\mb{a}+\mb{b}^\dagger\mb{b}+1\equiv \mb{H}$ close the ordinary $\mathfrak{sl}_2$ algebra, for which disentangling identities are known to reproduce \cite{PhysRevD.31.1988}
\[
\bra{0}\e^{x(\mb{F}+\mb{H}+\mb{E})}=\e^{-\ln(1-x)}\bra{0}\e^{\frac{x }{1-x}\mb{F}}.
\]
The less elegant one is based on brute force computation. In this case we can expand the exponential
\[
\e^{x(\mb{a}+\mb{b}^\dagger)(\mb{a}^\dagger+\mb{b})}=\sum_{n\geq 0}\frac{x^n}{n!}\sum_{k,j=0}^{n}\binom{n}{k}\bigl(\mb{b}^\dagger\bigr)^k \mb{a}^{n-k}\binom{n}{j}\bigl(\mb{a}^\dagger\bigr)^j\mb{b}^{n-j},
\]
and use the action of the oscillators on the Fock states, namely
\[
\bra{n}\bigl(\mb{a}^\dagger\bigr)^m= \sqrt{n(n-1)\cdots(n-m+1)}\bra{n-m},\qquad
\bra{n}\mb{a}^m= \sqrt{(n+1)\cdots(n+m)}\bra{n+m},
\]
to obtain
\[
\bra{0}\e^{x(\mb{a}+\mb{b}^\dagger)(\mb{a}^\dagger+\mb{b})}=\bra{0}\sum_{n,j\geq 0}x^j \binom{n+j}{j} \frac{(x\mb{a}\mb{b})^n}{n!}.
\]
The sum over $j$ (operator-independent part) is a known generating function ($|x|<1$)
\[
\sum_{j\geq 0}x^j \binom{n+j}{j}=\frac{1}{(1-x)^{n+1}}.
\]
 Finally, using
 \smash{$
\e^{\mb{a}\mb{b}}\mb{a}^\dagger\e^{-\mb{a}\mb{b}}=\mb{a}^\dagger+\mb{b}$}, \smash{$ \e^{\mb{a}\mb{b}}\mb{b}^\dagger\e^{-\mb{a}\mb{b}}=\mb{b}^\dagger+\mb{a}$},
we deduce the formula
\[
Z=\e^{-\ln(1-x)}\bra{0}\mb{O}\left(\frac{\mb{a}}{1-x}+\mb{b}^\dagger,\mb{a}^\dagger-\frac{\mb{b}}{1-x^{-1}}\right)\ket{0}.
\]

This derivation can be generalized to the case of multiple independent oscillators, both in the discrete case and in the continuous case whenever a suitable regularization scheme is employed to make sense of continuous products and singular integrals (e.g., via discretization, multiplicative integrals and deformed contours). We refer to \cite{Jimbo:1996ev} for a discussion in a related context.

In our application to the $\vep$-deformation, the regularization via discretization, namely $\dashint \d k\to \hbar\sum_{n\neq 0}$ and $\mb{a}(k)\to \mb{a}_n/\hbar$ (see also \cite{Lukyanov:1993pn}), brings us back to the $\q$-deformation recalled in Section~\ref{sec:vepAlg}. In that case, as shown in \cite{Lodin:2017lrc,Nieri:2015dts}, torus correlators computed via Clavelli--Shapiro formula can be written in terms of elliptic Gamma functions~($\Gamma_e$), which in the continuous limit $\hbar\to 0$ reduce to hyperbolic Gamma functions (i.e., double sine functions~$S_2$) thanks to the known asymptotic identity
\[
\Gamma_e\bigl(\e^{2\pi\i r x}|\e^{2\pi\i r \omega_1},\e^{2\pi\i r \omega_2}\bigr)\stackrel{r\to 0}{\simeq} \e^{-\frac{\i\pi}{12 r \omega_1\omega_2}(2x-\omega_1-\omega_2)}S_2(x|\omega_1,\omega_2)^{-1}.
\]
Up to zero mode contributions (which need to be taken care of separately) and divergent constant prefactors (which cancel in normalized correlators), this is a double check that (\ref{eq:qWcompact}) (and therefore the contour prescription we used) is the correct result.

\subsection*{Acknowledgements}

It is a pleasure to thank J.-E.~Bourgine, D.~Fioravanti, T.~Kimura, T.~Proch\'azka and Y.~Zenkevich for fruitful discussions. The author is deeply grateful to the Galileo Galilei Institute for Theoretical Physics for the hospitality during the 2024 Summer Workshop ``BPS Dynamics and Quantum Mathematics'' and the INFN for partial support during the early stages of this work. This work is also partially supported by the CRT Foundation under the grant 108399/2024.0434. The author also thank the anonymous referees who contributed to improve the final version of the paper.

\pdfbookmark[1]{References}{ref}
\LastPageEnding

\end{document}